\documentclass[12pt]{iopart}
\expandafter\let\csname equation*\endcsname\relax
\expandafter\let\csname endequation*\endcsname\relax
\usepackage{amssymb}
\usepackage{amsmath}
\usepackage{bbold}
\usepackage{graphicx}

\begin{document}

\title[Optical response induced by bound states in the continuum]{Optical response induced by bound states in the continuum in arrays of dielectric spheres}

\author{Evgeny N. Bulgakov$^{1,2}$ \& Dmitrii N. Maksimov$^{1,2}$}

\address{$^1$Reshetnev Siberian State University of Science and Technology, 660037, Krasnoyarsk,
Russia\\
$^2$Kirensky Institute of Physics, Federal Research Center KSC SB
RAS, 660036, Krasnoyarsk, Russia}
\ead{mdn@tnp.krasn.ru}
\vspace{10pt}
\begin{indented}
\item[] April 2018
\date{\today}
\end{indented}

\begin{abstract}
We consider optical response induced by bound states in the continuum (BICs) in arrays of dielectric spheres.
By combining quasi-mode expansion technique with coupled mode theory (CMT) we put forward a theory of the optical response
by high-Q resonance surrounding BICs in momentum space. The central result are
analytical expressions for the CMT parameters, which can be easily calculated from the eigenfrequencies and eigenvectors
of the interaction matrix of the scattering systems. The results obtained are verified in comparison against exact
numerical solutions to demonstrate that the CMT approximation is capable of reproducing Fano features in the spectral vicinity of
the BIC. Based on the quasi-mode expansion technique we derived the asymptotic scaling law for the CMT parameters in the vicinity of
the $\Gamma$-point. It is rigourously demonstrated that the line width in the CMT approximation exhibits different asymptotic behavior
depending on the symmetry of the BIC.
\end{abstract}

%
\vspace{2pc}

\section{Introductuion}
The phenomenon of light localization is of paramount importance in modern science and technology \cite{John12}.
One of the physical phenomena leading to the localization of light is optical bound states in the continuum (BICs) \cite{Hsu16}. BICs are
localized eigenmodes of Maxwell's equation embedded into the continuous spectrum of the scattering states, i.e. source-free
solutions which do not radiate into the far-zone albeit outgoing waves with the same frequency and wave vector are allowed
in the surrounding medium. In a view of time time reversal symmetry this implies that an ideal BIC is invisible from the far-zone
since it is also decoupled from any incident wave impinging onto the BIC supporting structure. At first glance this renders BICs totally
useless for practical purpose. However, if the scattering problem is granted an extra dimension by introducing a control parameter,
one can immediately see that the traces of BICs emerge in the scattering spectrum as narrow Fano features once the control parameter
is detuned from the BIC point \cite{Shipman2005,Foley14,Foley15,Cui16,Blanchard16}. The collapsing Fano feature is generally seen as a precursor of BIC not
only in optics but in
the fields of acoustics \cite{Hein12} and quantum mechanics \cite{Kim}. The emergence of Fano resonance is associated with critical light
enhancement \cite{Mocella15,Yoon15} which paves a way to important applications for light matter interaction including lasing \cite{Kodigala17,Bahari17,Ha18},
harmonics generation \cite{Ndangali2013, Wang18}, and bio-sensing \cite{Romano18}.

Among various set-ups BICs are known to exist in periodic dielectric structures
\cite{Venakides03,Marinica08,WeiHsu2013b,Monticone14,Yang2014,Bulgakov2015,Ni16,Sadrieva17,Monticone17}.
In that case any eigenmode of Maxwell's equations
is characterized by its Bloch wave number $\beta$ with respect to a certain axis of periodicity.
Thus, the periodicity by itself quite naturally offers the wave number as a control parameter for
optical response. Two classes of eigenmodes are generally discriminated in periodic structures.
The eigenmodes with frequency below the line of light $\omega=c\beta$ are always localized due to the total internal reflection. In contrast
the eigenmodes with frequency above the line of light are normally leaky, i.e. radiate to the outer space \cite{Monticone15}. In that context, BIC can be
seen as exceptional points of the leaky zones in which the far-field radiation from the leaky mode vanishes due to intricate destructive interference
between waves radiated from the infinite number of the elementary cells of the periodic structure. The BICs are observed from the dispersion
of the leaky-zone $Q$-factor as the points where the $Q$-factor diverges to infinity. This implies that spectrally any BIC is surrounded by
a family by high-$Q$ leaky modes \cite{Yuan17,Yuan18} with the $Q$-factort infinitely increasing as $\beta$ is tuned to the BIC point.

In this paper we propose a theory of optical response from such high-$Q$ leaky modes surrounding BICs in linear periodic arrays of dielectric spheres
\cite{Bulgakov2015}. To construct the theory of the resonant response two approaches are merged together. The first one is a rigorous quasi-mode expansion
technique which relies on the
bio-orthogonal basis of the interaction matrix of the scattering system \cite{Bai13,Armitage14,Vial14,Vial16,Alpeggiani17,Fehrembach18}. The second approach is coupled mode theory (CMT) \cite{Fan03,Suh04,Hamam07}
which proved itself
as an efficient tool for approximating the resonant response of optical systems. In what follows we shall establish a link between the two approaches to provide a clear physical
picture in terms of the leaky mode coupling coefficients leading to simple expressions for amplitudes of the scattered waves. We shall reveal the asymptotic behavior
of the coupling coefficients in the spectral vicinity of BIC and numerically demonstrate the validity of the proposed approach.

\section{Scattering theory}

\begin{figure}[t]
\includegraphics[width=0.8\textwidth,trim={0.0cm 1.9cm 0cm 0cm},clip]{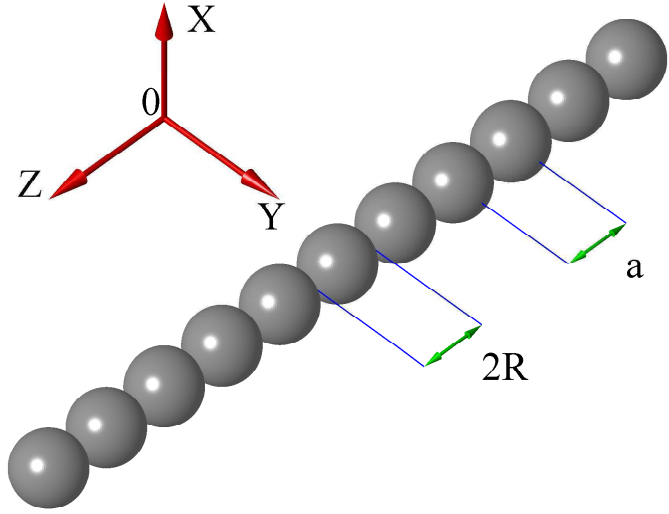}
\caption{(Color online) Array of dielectric spheres of radius $R$ periodically arranged along the $z$-axis with period $a$. }
\label{fig1}
\end{figure}

The system under consideration is an array of dielectric spheres with permittivity $\epsilon=15$ in air shown in Fig \ref{fig1}.
The electromagnetic (EM) field can be found from time stationary Maxwell's equations
\begin{eqnarray}
\nabla\times {\bf E}= i k_0 {\bf H}, \notag \\
\nabla\times {\bf H}= -i k_0 \epsilon({\bf r}){\bf E}, \label{Maxwell}
\end{eqnarray}
where {\bf H}, {\bf E} are the magnetic and electric vectors, $\epsilon({\bf r})$ is the dielectric permittivity, and
$k_0$ is the vacuum wave number, $k_0=\omega/c$ with $\omega$ as the frequency and $c$ as the speed of light. In what follows
we set $c=1$ to measure the frequency in the units of distance.
The solution for the EM field scattered by the array could be found by using the
method developed by Linton, Zalipaev, and Thompson \cite{Linton13}.
According to \cite{Linton13} both $\bf{E}$ and $\bf{H}$ outside the spheres
are written as a series over
spherical vector harmonics ${\bf N}^{m}_{l}({\bf r}),{\bf M}^{m}_{l}({\bf r})$  in the following form  
\begin{eqnarray}\label{solution_expansion}
{\bf E}({\bf r})=\sum^{\infty}_{j=-\infty}e^{iaj \beta}\sum^{\infty}_{l=m^{*}}
\left[\overline{a}_l^m{\bf M}^{m}_{l}({\bf r}_j)+\overline{b}_l^m{\bf N}^{m}_{l}({\bf r}_j)\right] \notag\\
{\bf H}({\bf r})=-i\sqrt{\epsilon}\sum^{\infty}_{j=-\infty}e^{iaj \beta}\sum^{\infty}_{l=m^{*}}
\left[\overline{a}_l^m{\bf N}^{m}_{l}({\bf r}_j)+\overline{b}_l^m{\bf M}^{m}_{l}({\bf r}_j)\right],
\end{eqnarray}
where $j$ is the number of the sphere in the array, $a$ is the distance between the centers of the spheres, ${\bf r}_j$ - coordinate vector in the $j_{\mathrm{th}}$
sphere reference frame, and $m^*=max(1,m)$.
When the scattering problem is addressed the EM field is found
by solving a set of linear equations \cite{Bulgakov2015}
\begin{equation}\label{matrix}
\widehat{\cal{L}}\bf{p}=\bf{q},
\end{equation}
where $\bf{p}$ is the vector of the expansion coefficients of the scattered EM field
\begin{equation}
{\bf p}=\left\{\overline{a}^m_l, \overline{b}^m_l\right\},
\end{equation}
${\bf q}$ is the vector of the expansion coefficients of the incident wave
\begin{equation}\label{q}
{\bf q}=\left\{c_l^m, d_l^m\right\},
\end{equation}
and ${\cal \widehat{L}}$ is the interaction matrix of the scattering system. We spare the reader of the exact, rather cumbersome, expression
for ${\cal \widehat{L}}$ since it is already given in \cite{Bulgakov2015} (and implicitly in the original paper \cite{Linton13}).

Given that matrix ${\cal \widehat{L}}$ is a known quantity we yet have to define vector $\bf{q}$. The expression for $\bf{q}$ depends on the configuration
of the incident EM field.  The most obvious choice for probing the resonant response is a monochromatic plane wave.
The plane wave is specified by its polarization;
a transverse electric (TE) wave has its electric vector orthogonal to the array axis, while for a transverse magnetic (TM) wave it
is the magnetic vector that is orthogonal to the axis of the array.
Importantly, the BICs in arrays of dielectric spheres always have quantized orbital angular momentum (OAM) $m$ \cite{Bulgakov2015}, which reflects
the rotational symmetry of the system.
Since OAM is preserved by scattering of the cylindrical waves, the scattering problem for a plane wave can be solved independently for each $m$.
In what follows we shall only consider scattering in subspace with OAM equal to that of the BIC having in mind that the response
to a plane wave could be found relaying on the expansion of a plane wave into cylindrical Bessel functions \cite{Kong00}. As it was demonstrated
in \cite{Bulgakov2015} the sharp resonant feature in the spectral vicinity of a BIC is clearly visible in the total scattering
cross-section since subspaces with OAM different from that of the BIC provide only smooth non-resonant background contribution.
For each $m$ the incident TM (TE) wave is defined as
\begin{equation}\label{Bessel_channel}
E^{(m)}_z \left(H^{(m)}_z\right)=\frac{1}{\sqrt{C({\bf k})}}\exp(im\phi+ik_zz)J_{m}(k_{\perp}\rho)
\end{equation}
with
\begin{equation}
k_{\perp}^2=k_0^2-k_z^2,
\end{equation}
where $\{\rho, \phi, z\}$ are the cylindrical coordinates, $k_z$ is the $z$ component of the wave vector,
 and $C({\bf k})$ is normalization constant to be specified later on in the text.

Similarly to Eq. (\ref{solution_expansion})
the mode shape of the incoming wave could be expanded into vector spherical harmonics
\begin{eqnarray}\label{expansion_channel}
{\bf E}^{({\rm inc})}({\bf r})=\sum^{\infty}_{j=-\infty}e^{iaj k_z}\sum^{\infty}_{l=m^{*}}
\left[{c}_l^m{\bf M}^{m}_{l}({\bf r}_j)+{d}_l^m{\bf N}^{m}_{l}({\bf r}_j)\right] \notag\\
{\bf H}^{({\rm inc})}({\bf r})=-i\sqrt{\epsilon}\sum^{\infty}_{j=-\infty}e^{iaj k_z}\sum^{\infty}_{l=m^{*}}
\left[{c}_l^m{\bf N}^{m}_{l}({\bf r}_j)+{d}_l^m{\bf M}^{m}_{l}({\bf r}_j)\right].
\end{eqnarray}
According to \cite{Kong00} the expansion coefficient of the TE-polarized incident wave can be written as
\begin{eqnarray}\label{cd_def}
{c}_l^m=\frac{i^mk_0}{\sqrt{C({\bf k})}k_{\perp}}F_{l,m}\pi_{l,m}(\theta), \notag\\
{d}_l^m=\frac{-i^mk_0}{\sqrt{C({\bf k})}k_{\perp}}F_{l,m}\tau_{l,m}(\theta), \label{TE_in}
\end{eqnarray}
where $\theta$ is the angle between the wave vector ${\bf k}$ and the array axis $z$, $\cos(\theta)=k_z/k_0$,
\begin{eqnarray}
\pi_{l,m}(\theta)=-\frac{\partial}{\partial \theta}P^m_l[\cos(\theta)], \notag \\
\tau_{l,m}(\theta)=\frac{m}{\sin(\theta)}P^m_l[\cos(\theta)] \label{pi_tau}
\end{eqnarray}
with $P^m_l$ as the associated Legendre polynomials,
and
\begin{equation}\label{Flm}
F_{l,m}=(-1)^mi^l\sqrt{\frac{4\pi(2l+1)(l-m)!}{(l+m)!}}.
\end{equation}
At the same time for the TM-polarized waves we have
\begin{eqnarray}
{c}_l^m=\frac{i^{m+1}k_0}{\sqrt{C({\bf k})}k_{\perp}}F_{l,m}\tau_{l,m}(\theta), \notag \\
{d}_l^m=-\frac{i^{m+1}k_0}{\sqrt{C({\bf k})}k_{\perp}}F_{l,m}\pi_{l,m}(\theta), \label{TM_in}
\end{eqnarray}

For a further analysis we will use the quasi-modal expansion \cite{Vial14} based on the biorthogonal
basis of the left ${\bf y}_n$ and right ${\bf x}_n$ right
eigenvectors of matrix $\widehat{{\cal L}}$
\begin{equation}\label{basis}
\widehat{{\cal L}}{\bf x}_n=\lambda_n {\bf x}_n, \
\widehat{{\cal L}}^{\dagger}{\bf y}_n=\lambda_n^* {\bf y}_n, \
\end{equation}
It should be noticed that $\widehat{{\cal L}}$, and consequently  $\lambda_s, {\bf x}_s, {\bf y}_s$,
are dependent on both $\beta$ and $k_0$ \cite{Bulgakov2015}. The biorthogonal eigenvectors obey the following normalization
condition
\begin{equation}\label{bionorm}
{\bf y}_n^{\dagger}{\bf x}_{n'}={\bf x}_n^{\dagger}{\bf y}_{n'}=\delta_{n,n'}.
\end{equation}
For further convenience the right eigenvector can be explicitly written as a vector of expansion coefficients over spherical harmonics
in Eq. (\ref{solution_expansion})
\begin{equation}\label{left}
{\bf y}=\left\{ {a}^m_l,  {b}^m_l\right\},
\end{equation}
where for simplicity we omitted subscript $n$. Substituting the coefficients ${a}^m_l,  {b}^m_l$ to Eq. (\ref{solution_expansion}) instead
 of $\overline{a}^m_l,  \overline{b}^m_l$ one can produce the profile of the optical quasi-mode.

Taking into account Eqs. (\ref{basis}, \ref{bionorm}) the inverse of $\widehat{\cal L}$ is given by
\begin{equation}\label{inverse}
\widehat{{\cal L}}^{-1}=\sum_n \frac{1}{\lambda_n} {\bf x}_n {\bf y}_n^{\dagger}.
\end{equation}
Applying Eq. (\ref{inverse})
we can write the solution of the scattering problem Eq. (\ref{matrix}) under illumination by the incident wave Eq. (\ref{Bessel_channel})
in the following form
\begin{equation}\label{solution}
{\bf p}^{(p)}=\sum_n \frac{w^{(p)}_{n}}{\lambda_n} {\bf x}_n
\end{equation}
with
\begin{equation}\label{coupling_strengh}
w^{(p)}_n={\bf y}_n^{\dagger}{\bf q}^{(p)},
\end{equation}
where $p=e,h$ is used for TM or TE waves, respectively. The entries of vector ${\bf q}^{(p)}=\{c_l^m, d_l^m \}$ are defined by Eqs. (\ref{TE_in}, \ref{TM_in})
depending on polarization.
Further on the quantity $w_n$ introduced in Eq. (\ref{coupling_strengh}) will
be refereed to as the {\it quasi-mode coupling strength}.
The quasi-mode coupling strength can be expressed through expansion coefficients $a_l^m, b_l^m$ as
\begin{eqnarray}\label{wh}
w^{(h)}=\frac{i^mk_0}{\sqrt{C({\bf k})}k_{\perp}}\sum_lF_{l,m}\left[\pi_{l,m}({a}^m_l)^* -\tau_{l,m}({b}^m_l)^*\right], \notag \\
w^{(e)}=\frac{i^{m+1}k_0}{\sqrt{C({\bf k})}k_{\perp}}\sum_lF_{l,m}\left[\tau_{l,m}({a}^m_l)^* -\pi_{l,m}({b}^m_l)^*\right], \label{we}
\end{eqnarray}
where we again omitted subscript $n$ for simplicity.

\section{Bound states in the continuum}

The linear arrays of dielectric spheres possesses rotational symmetry about the axis of periodicity, and, thus, preserve the orbital angular momentum (OAM) of
light. As consequence the system supports BICs with quantized OAM including twisted BICs with non-zero OAM \cite{Bulgakov16a,Bulgakov17a}. For simplicity
and brevity of presentation in this paper we
restrict ourself with the case of zero OAM $m=0$ \cite{Bulgakov2015}.
We shall also stay in the the domain where only the zeroth diffraction order is allowed in the far-field radiation
\begin{equation}\label{domain}
k_za<k_0d<2\pi-k_za
\end{equation}
This allows us to equate the Bloch wave number with the $z$-component of the incident wave vector,  $k_z=\beta$. Under condition Eq. (\ref{domain})
the system supports only two far-field scattering channels, one TM- and one TE-polarized.
BICs are source-free solutions of Eq. (\ref{matrix}) which exist even without the array being illuminated from the far zone
to yield a simple condition
\begin{equation} \label{condition}
\det[\widehat{\cal L}(k_0,\beta)]=0.
\end{equation}
The above condition means that in the BIC point there is an eigenvector ${\bf y}_0$ with zero eigenvalue
\begin{equation}
\widehat{\cal L}^{\dagger}{\bf y}_{0}=0
\end{equation}
obviously corresponding to the BIC mode shape. Technically, the BICs can be found by searching for the zeros of
$\det[\widehat{\cal L}(k_0,\beta)]$ in $k_0,\beta$-space. It is, however, more numerically efficient to only find the eigenvalue
of $\widehat{\cal L}$ with the least absolute value. With respect to $\beta$ the BICs can be split into two classes \cite{Bulgakov2015}, in-$\Gamma$ standing wave BICs with $\beta=0$ and
off-$\Gamma$ travelling wave (Bloch) BICs with $\beta \neq 0$. In this paper we shall be concerned with in-$\Gamma$ BICs, although the application
of the proposed theory to off-$\Gamma$ BICs is straightforward.

\begin{figure}[t]
\includegraphics[width=1\textwidth,trim={0.0cm 0cm 0cm 0cm},clip]{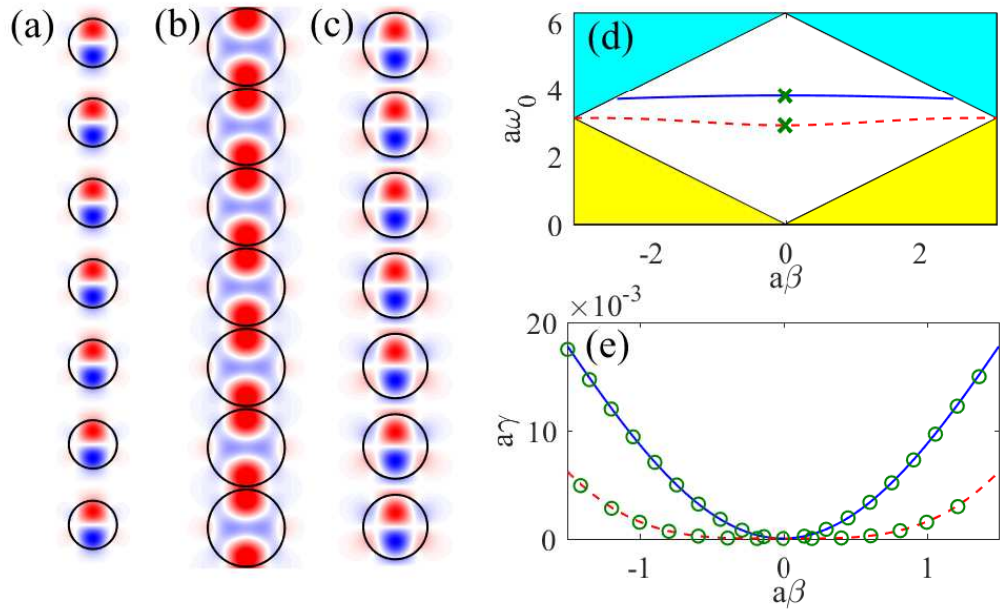}
\caption{(Color online) BICs in the dielectric array from Fig. \ref{fig1} with $\epsilon=15$.
(a) The $z$-component of magnetic field, $H_z$ for the symmetry protected TE
BIC with $ak_0=3.8141$ and $R=0.3000a$. (b) $H_z$
for the non-symmetry protected TE
BIC with $ak_0=2.9375$ and $R=0.480079a$. (c) The $z$-component of electric field, $E_z$ for symmetry protected
TM BIC with $ak_0=3.6021$ and $R=0.4000a$. (d) Dispersion of the real part of the eigenfrequency, $\omega_0$ for the TE leaky modes
hosting the BICs; solid blue - leaky mode with a symmetry protected BIC, dash red - leaky mode hosting the non-symmetry protected BIC. The white area
is the domain where two scattering channels are open Eq. (\ref{domain}). The positions of the in-$\Gamma$ BICs are marked
by crosses. (e) Dispersion of the imaginary part of the eigenfrequency, $\gamma$ for the TE leaky modes
in the vicinity of the $\Gamma$-point; solid blue - leaky mode with the symmetry protected BIC, dash red -
leaky mode with the non-symmetry protected BIC. Green circles show the results obtained with Eq. (\ref{Gamma}). }
\label{fig2}
\end{figure}
As it was mentioned in the introduction the BICs are exceptional point of the leaky-zone in which the $Q$-factor diverges to infinity. Each leaky zone is
characterized by dispersion of the complex eigenfrequency
\begin{equation}\label{complex_eig}
\Omega=\omega_0-i\gamma,
\end{equation}
where $\omega_0$ and $\gamma$ are the position and the width of the resonance, respectively.
The Q-factor is defined as
\begin{equation}
Q=\frac{\omega_0}{2\gamma}
\end{equation}
with $\gamma$ obviously vanishing in the point of a BIC.
The complex eigenfrequencies can be found by analytic extension of Eq. (\ref{condition}) to the complex plane as function of $k_0$.
In Fig. \ref{fig2} we show results of numerical simulations for both TE-and TM-polarized waves with $m=0$. Importantly in case $m=0$ the
waves of TE and TM polarizations are not coupled by matrix $\widehat{\cal{L}}$ \cite{Bulgakov2015}. Thus, all waves are pure TE- or TM-polarized
and only a single decay channel is allowed
for each polarization. In Fig. \ref{fig2}. (a, b) we show the mode profiles
of TE BICs. Notice, that in Fig. \ref{fig2} (a) $H_z$ is antisymmetric with respect to the axis of the array. That is an example of so-called symmetry protected
BICs which are symmetrically mismatched with the TE scattering channel. In contrast to the previous case the BIC from Fig. \ref{fig2} (b) is not symmetry protected.
The mode profile of a TM symmetry protected BIC is shown in Fig. \ref{fig2} (c).
In Fig. \ref{fig2} (d, e) we show the dispersion of $\omega_0$ and $\gamma$, respectively for TE-leaky modes. Notice that on approach to the $\Gamma$-point
for the non-symmetry protected BIC $\gamma$ vanishes faster than for the symmetry protected ones.
As it was found in \cite{Bulgakov17oe} this reflects in different asymtotics of the $Q$-factor for symmetry protected and non-symmetry protected BICs at
the $\Gamma$-point, the difference to be explained in Sec. \ref{Sec5}.

\section{Coupled mode theory}
To understand the features of the resonant response in the spectral vicinity of a BIC in
infinite arrays we resort to coupled mode theory (CMT) \cite{Fan03,Suh04}. CMT is a rather generic phenomenological approach
relying on the modal representation of the EM field in the scattering domain. Each mode is viewed as
an environment coupled oscillator with complex eigenfrequency Eq. (\ref{complex_eig}).
One condition essential for CMT is the energy conservation. To account for the energy conservation the system's modes must
be energy normalized. Moreover,
the mode shapes of the scattering channels must be normalized to supply (retrieve) a unit energy flux into (from) the scattering domain.
To explicitly define the channels
we consider the total EM field far from the scattering domain. Following \cite{Ruan12} we decompose the far-field into incoming and outgoing
waves
\begin{equation}\label{total_1}
E_z^{\rm (tot)}\left(H_z^{\rm (tot)}\right)=\frac{1}{\sqrt{C({\bf k})}}\left[a_{e,(h)}^{(+)}H^{(2)}_{m}
(k_{\perp}\rho)+a_{e,(h)}^{(-)}H^{(1)}_{m}(k_{\perp}\rho) \right]
e^{im\phi+ ik_z z},
\end{equation}
where $H^{(1,2)}_{m}(x)$ are the Hankel functions, $a_{e,h}^{(+)}$ is the amplitude of the incoming wave,
and $a_{e,h}^{(-)}$ is the amplitude of the outgoing wave. The vectors of the incoming ${\bf a}^{(+)}=\{a_{e}^{(+)}, a_{h}^{(+)} \}$
and the outgoing ${\bf a}^{(-)}=\{a_{e}^{(-)}, a_{h}^{(-)} \}$ amplitudes are linked through the $\widehat{\cal S}$-matrix
\begin{equation}\label{S}
{\bf a}^{(-)}=\widehat{\cal S}{\bf a}^{(+)}.
\end{equation}
The $\widehat{\cal S}$-matrix defined in Eq. (\ref{S}) has the following property
\begin{equation} \label{cond4}
\widehat{\cal S}^T=\widehat{\sigma}_z\widehat{\cal S}\widehat{\sigma}_z,
\end{equation}
where $\widehat{\sigma}_z$ is the third Pauli matrix.
Eq. (\ref{cond4}) can be easily proved using $\widehat{\cal S}$-matrix unitarity taking  and into account that that the magnetic field
is a quasi-vector which flips its sign under the time reversal operation. At this point we remind the reader that for $m=0$
the polarization conversion is forbidden and, hence, $\widehat{\cal S}$ is diagonal.
On the other hand, since the incident wave Eq. (\ref{Bessel_channel}) is defined trough the Bessel rather than the Hankel
functions we can write
\begin{equation}\label{total_2}
E_z^{(tot)}\left(H_z^{(tot)}\right)=\frac{1}{\sqrt{C({\bf k})}}\left[a_{e,(h)}^{(+)}2J_{m}
(k_{\perp}\rho)+a_{e,(h)}H^{(1)}_{m}(k_{\perp}\rho) \right]
e^{im\phi+ ik_z z},
\end{equation}
where $a_{e,h}$ are unknown coefficients to be found by solving the scattering problem. The far-field solution
can be expressed through matrix $\widehat{\cal T}$ as
\begin{equation}\label{T}
{\bf a}=\widehat{\cal T}{\bf a}^{(+)},
\end{equation}
where ${\bf a}=\{a_{e}, a_{h} \}$. For the further convenience we designate the elements of $\widehat{\cal T}$ as follows
\begin{equation}\label{Tdef}
\widehat{\cal T}=
\left\{
\begin{array}{cc}
 t_{e,e} & 0 \\
0 & t_{h,h}
\end{array}
\right\}.
\end{equation}
Following \cite{Ruan12} with the use of the identity $2J_{m}(x)=H^{(1)}_{m}(x)+H^{(2)}_{m}(x)$ one finds
from Eqs. (\ref{total_1}, \ref{total_2}) that
\begin{equation}\label{TS}
\widehat{\cal T}=\widehat{\cal S}-\widehat{\mathbb{1}}.
\end{equation}
Notice that unlike $\widehat{\cal S}$ matrix $\widehat{\cal T}$ is generally non-unitary.

The
goal of this paper is to construct a CMT for matrix $\widehat{\cal S}$. The mode shape of the TM (TE) scattering channel
are implicitly defined by Eq. (\ref{total_1}) as
\begin{equation}\label{channel}
E_z^{\rm (inc, out)}\left(H_z^{\rm (inc, out)}\right)=\frac{1}{\sqrt{C({\bf k})}}H^{(1,2)}_{m}
(k_{\perp}\rho)
e^{im\phi\pm k_z z},
\end{equation}
where $\rm (inc, out)$ stand for incoming and outgoing wave, respectively.
By requiring that each scattering cannel supplies a unit energy flux
per period of the array we can find the normalization constant in Eq. (\ref{channel}) as
\begin{equation}\label{norm}
C({\bf k})=\frac{2k_0}{k^2_{\perp}}.
\end{equation}
Technically, the above result is obtained by finding the total Poynting of the channel function Eq. (\ref{channel}) through the cylindrical
surface enveloping the elementary cell of the array.
Under
above condition the response of a {\it single} resonant mode to impinging waves is described by a following equation \cite{Ruan12}
\begin{eqnarray}\label{CMT1}
b=\frac{{\boldsymbol \kappa}^{T}{\bf a}^{(+)}}{i(\omega_0-\omega)+\gamma}, \\ \label{CMT2}
{\bf a}^{(-)}=\widehat{\cal C}{\bf a}^{(+)}+b{\bf d},
\end{eqnarray}
where $b$ is the amplitude of the resonant mode $\widehat{\cal C}$ is the scattering matrix
of direct (non-resonant) transmission path, and vectors ${\boldsymbol \kappa}, {\bf d}$ describe the coupling between the resonant mode
and the incoming and outgoing waves, respectively.
Most importantly the
quantities $\gamma$, ${\bf d}$, and $\widehat{\cal C}$ introduced in Eqs. (\ref{CMT1}, \ref{CMT2})
are not independent. The relationships between $\gamma$, ${\bf d}$, and $\widehat{\cal C}$ can
be established by performing the time reversal operation \cite{Fan03,Suh04,Ruan12}. To apply the time reversal arguments
one has to take into account that under the time reversal the cannel functions Eq. (\ref{channel}) change the sign of both $\beta$
and $m$. This has been rigourously done in \cite{Ruan12,Guo17}. In our case, however, the system possesses both rotational
symmetry and symmetry with respect to the mirror reflection in the $x0y$-plane. Thus we can conclude that $\gamma$, ${\bf d}$, and $\widehat{\cal C}$
are immune to the time reversal. Then, following the standard procedure from \cite{Fan03} one finds
\begin{eqnarray}\label{cond1}
\widehat{\cal C}\widehat{\sigma}_z{\bf d}^{*}=-{\bf d}, \\ \label{cond2}
{\bf d}^{\dag}{\bf d}=2\gamma, \\ \label{cond3}
{\boldsymbol \kappa}= \widehat{\sigma}_z{\bf d}.
\end{eqnarray}
Notice that the above expression only differ from the standard two channel CMT \cite{Fan03} by emergence of $\widehat{\sigma}_z$
which simply reflects the behavior of the channel functions Eq. (\ref{channel})  under the time reversal.
The final expression for the scattering matrix reads
\begin{equation}\label{CMT_S_matrix}
\widehat{\cal S}=\widehat{\cal C}+\frac{{\bf d}{\bf d}^{T}}{i(\omega_0-\omega)+\gamma}\widehat{\sigma}_z.
\end{equation}
Now using Eqs. (\ref{TS}, \ref{CMT_S_matrix}) one can write the solution under illumination by the Bessel waves
\begin{equation}\label{CMT_solution}
{\bf a}=\left(-\widehat{\mathbb{1}}+\widehat{\cal C}+\frac{{\bf d}{\bf d}^{T}}{i(\omega_0-\omega)+\gamma}\widehat{\sigma}_z\right) {\bf a}^{(+)}.
\end{equation}

 To establish a link between Eq. (\ref{solution}) and the CMT solution Eqs. (\ref{CMT_solution})
we consider the resonant contribution to Eq. (\ref{solution}), i.e  the singular term that with
vanishing $\lambda_n$ in the denominator. In what follows
we omit the subscript $n$ having in mind that only the resonant contribution is considered. 
The resonant term reads
\begin{equation}\label{resonant}
{\bf f}=\frac{{\bf w}^T {\bf a}^{(+)}}{\lambda}{\bf {x}},
\end{equation}
where ${\bf w}=\left\{ w^{(e)}, w^{(h)}\right\} $.
Let us now write a series expansion in $\omega$ in the vicinity of the
leaky mode resonant eigenfrequency $\omega_0$ at with a fixed $\beta$ near the normal incidence
\begin{eqnarray}\label{exp_def}
{\bf w}= {\bf w}_0+{\bf w}_1(\omega-\omega_0), \
\lambda= \lambda_0+\lambda_1(\omega-\omega_0), \
{\bf x}= {\bf x}_0+{\bf x}_1(\omega-\omega_0),
\end{eqnarray}
where
\begin{equation}
{\bf w}_{0,1}=\left\{ w^{(e)}_{0,1}, w^{(h)}_{0,1}\right\}.
\end{equation}
Notice, that all quantities introduced in Eq. (\ref{exp_def}) are dependent on $\beta$.
Expanding Eq. (\ref{resonant}) up to the first order in $\omega-\omega_0$ we have
\begin{equation}\label{expansion}
{\bf f}=\left({{\bf g}_0^T{\bf a}^{(+)}}\right){\bf x}_0 +\left({\bf g}_1^T{\bf a}^{(+)}\right){\bf x}_1
\end{equation}
with
\begin{eqnarray}\label{expansion_1}
{\bf g}_0=\frac{\left( {\bf w}_0 -\frac{\lambda_0}{\lambda_1}{\bf w}_1 \right)}{\lambda_0+\lambda_1(\omega-\omega_0)}
 +\frac{1}{\lambda_1}{\bf w}_1,\\ \label{expansion_2}
{\bf g}_1=\frac{-\frac{\lambda_0}{\lambda_1}{\bf w}_0}{\lambda_0+\lambda_1(\omega-\omega_0)}
 +\frac{1}{\lambda_1}{\bf w}_0.
\end{eqnarray}
Before comparing Eq. (\ref{expansion}) against Eq. (\ref{CMT1}) we have to renormalize the resonant
mode to support a unit energy per period of the array
\begin{equation}\label{norm_x}
\ {\bf \overline{x}}_0={\bf x}_0/\sqrt{{C}_0}
\end{equation}
where ${C}_0$ is the normalization constants. Since the optical quasi-mode ${\bf \overline{x}}_0$ is unit-normalized
the outgoing filed can be produced with the CMT coupling vector ${\bf d}$
\begin{equation}\label{resonant_solution}
{\bf a}=\sqrt{C_0}\left({\bf d}{\bf g}_0^T\right){\bf a}^{(+)}.
\end{equation}

\begin{figure}[t]
\includegraphics[width=1\textwidth,trim={0.0cm 0cm 0cm 0cm},clip]{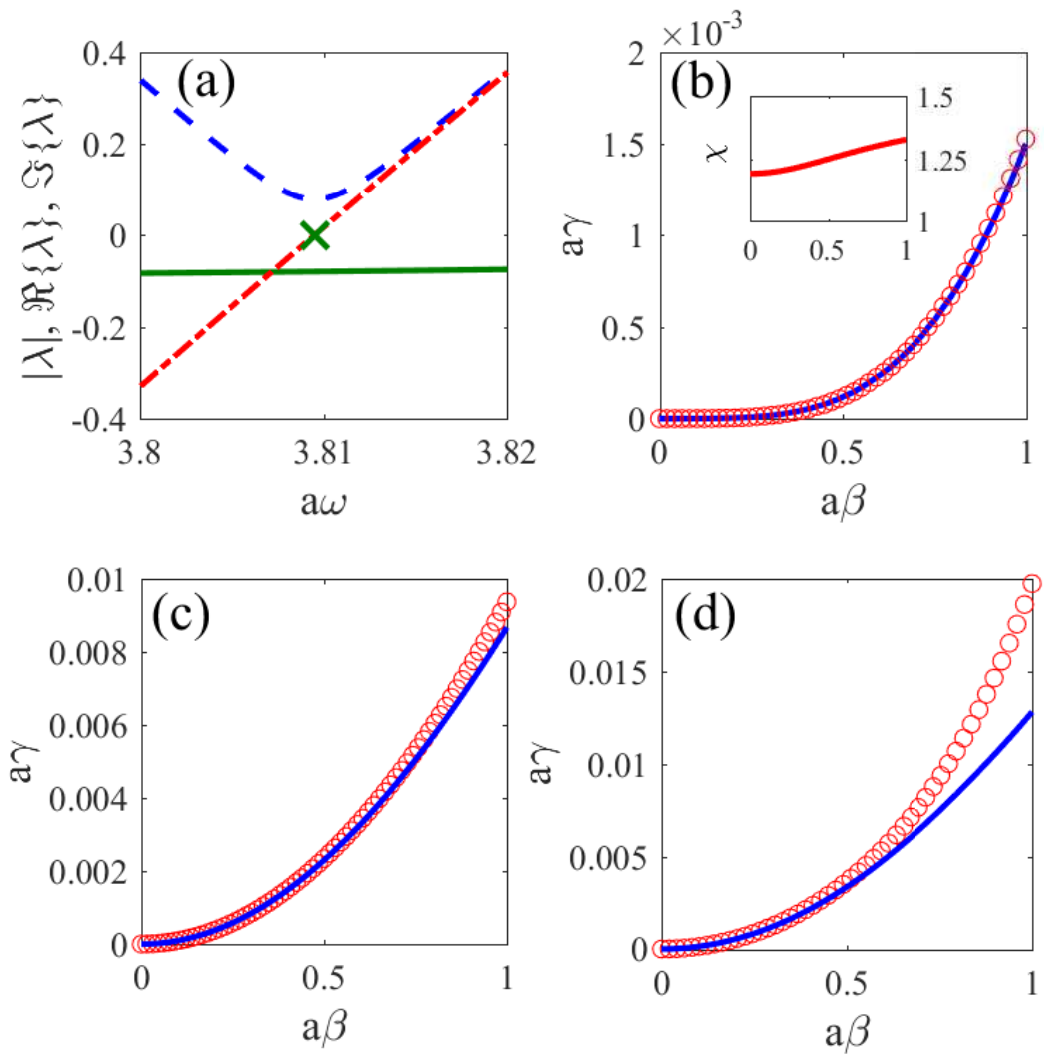}
\caption{(Color online) (a) Resonant eigenvalue $\lambda$ in the spectral vicinity $a\beta=0.5$
 of the symmetry protected BIC from Fig. \ref{fig2} (a); dash blue line - $|\lambda|$, solid green line - $\Re\{\lambda\}$, dash-dot red line
 $\Im\{\lambda\}$; the position of the resonance is shown by green cross.
 (b) Resonant width $\gamma$ found from Eq. (\ref{Gamma}) for the leaky-zone with the non-symmetry protected BIC \ref{fig2} (b) - solid blue line, and as $\gamma=|\gamma_h|^2/2$ from Eq. (\ref{final}). The inset shows
 the dependance of the phase of the CMT coupling coefficient $\chi$ on $\beta$.
 (c,d) The same as in (b) for TE and TM symmetry protected BIC from Fig. \ref{fig2} (a) and Fig. \ref{fig2} (c), respectively.}
\label{fig3}
\end{figure}

Equating the resonant term in Eq. (\ref{CMT_solution}) to that in Eq. (\ref{resonant_solution})
we obtain the following equations
\begin{eqnarray}\label{equations} \nonumber
{-i\sqrt{C_0}}\left({\bf w}_0 -\frac{\lambda_0}{\lambda_1}{\bf w}_1  \right)=\lambda_1\widehat{\sigma}_z{\bf d},\\
{\sqrt{C_0}\gamma}\left({\bf w}_0 -\frac{\lambda_0}{\lambda_1}{\bf w}_1  \right)=\lambda_0\widehat{\sigma}_z{\bf d}.
\end{eqnarray}
One can immediately see from the above equations that
\begin{equation}\label{Gamma}
\gamma=-i\frac{\lambda_0}{\lambda_1}.
\end{equation}
The coefficient $\lambda_1$ is a non-vanishing quantity, hence $\lambda_0$ amd $\gamma$ exhibit
the same asymptotics in the vicinity of the BIC. Notice, that the resonant term in Eq. (\ref{expansion_2})
can be dropped because it vanishes with $\lambda_0$. Eq. (\ref{Gamma}) can be used for finding the resonant width of the high-Q leaky zone without
the use of analytic extension to the complex plane. One can see from Fig. \ref{fig2} (e) that Eq. (\ref{Gamma}) demonstrates a good agreement with
the true resonant width found from analytic extension in the vicinity of the $\Gamma$-point. For example, if $a\beta=0.5$ for the leaky mode with the symmetry protected BIC we have
$a\Omega=3.80950-i0.00231$, while the analytic extension yields $a\Omega=3.80947-i0.002299$.
Remarkably, by definition Eq. (\ref{exp_def}) it is not guaranteed that
$\gamma$ found from Eq. (\ref{Gamma}) is a real positive quantity.
It appears to be a challenging task to prove that (\ref{Gamma}) is generally real positive.
Our numerical results, however, indicate that it is always real positive
in consistence with definition Eq. (\ref{complex_eig}). Moreover, it is found that $\lambda_0$ is always real negative, while $\lambda_1$
is imaginary positive.
In Fig. \ref{fig3} (a) we show the dependance of $\lambda$ on $\omega$ in the spectral vicinity of the symmetry protected BIC from Fig. \ref{fig2} (a).
Next, we further analyze Eqs. (\ref{equations}) depending
on the polarization of the scattering channels for $m=0$.

\begin{figure}[t]
\includegraphics[width=1\textwidth,trim={0.0cm 4cm 0cm 4cm},clip]{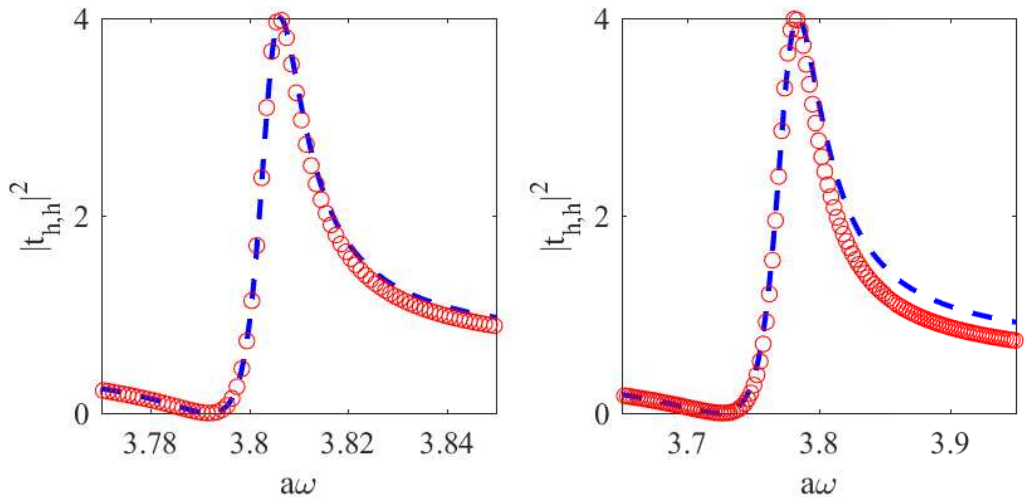}
\caption{(Color online) Optical response induced by the symmetry protected BIC from Fig. \ref{fig1} (a).
Blue dash line - exact numerical solution by Eq. (\ref{matrix}), red circles - CMT approximation. Left panel - $a\beta=0.7424$, right panel - $a\beta=1.5000$. }
\label{fig4}
\end{figure}

\subsection{TE-modes}
Using Eq. (\ref{cond2}) vector ${\bf d}$ for TE modes can be written as
\begin{equation}\label{d_vector}
{\bf d}=
\left\{
\begin{array}{l}
0 \\
\gamma_h
\end{array}
\right\}, \ 2\gamma=|\gamma_h|^2
\end{equation}
One can see from Eqs (\ref{Gamma}, \ref{d_vector}) that
$\lambda_0$ vanishes faster then $\gamma_h$, and
we can drop the second term in the round brackets in Eq. (\ref{equations}). For the
CMT coupling coefficient we have
\begin{equation}\label{gamma}
\gamma_h=i\frac{\sqrt{C_0}}{\lambda_1}w^{(h)}_{0}
\end{equation}
which dictates the same asymptotic behavior of $\gamma_h$ and $w^{(0)}_h$.
By applying Eq. (\ref{cond3}) we also have
\begin{equation}
{C_0}=2\gamma\left|\frac{\lambda_1}{w^{(h)}_{0}}\right|^2.
\end{equation}
Thus, we can express all features of the resonant response through
only three parameters $\lambda_0$, $\lambda_1$, and $w_0$.
Taking into account the $\lambda_1$ is real positive we can write
\begin{equation}\label{nophase}
\gamma_h=\sqrt{2\gamma} \ {\rm arg}\left(w^{(h)}_{0}\right).
\end{equation}
Unfortunately, Eq. (\ref{nophase}) can not be unambiguously applied for finding $\gamma_h$ because the biorthogonal
eigenvectors Eq. (\ref{bionorm}) and, consequently, the quasi-mode coupling strength $w^{(h)}$ are defined up to an
arbitrary phase factor.
To avoid the problem with the arbitrary phase factor we apply
an alternative approach for calculating $\gamma_h$ based on finding the far-field structure of quasi-mode ${\bf x}$.
The details of the approach are described in \ref{appendix}. The final result reads
\begin{equation}\label{final}
\gamma_h=\frac{\pm k_{\perp}}{2} \sqrt{\frac{1}{|\lambda_1| a k_0^3}}\overline{w}^{(h)}_{0}.
\end{equation}
Notice that Eq. (\ref{final}) allows to determine $\gamma_h$ up to its sign. This, however, does not affect the reflection coefficient defined
by Eq. (\ref{CMT_solution}). In Fig. \ref{fig3} (b,c) we show the resonant width $\gamma=|\gamma_h|^2/2$ found from Eq. (\ref{final}) for the leaky zones with
the non-symmetry protected and symmetry protected TE BICs, respectively, in comparison
against Eq. (\ref{Gamma}) to demonstrate the agreement between the two approaches in the vicinity of the $\Gamma$-point. At large $|\beta|$, however,
$\gamma$ found from Eq. (\ref{final}) deviates from the true resonant width. That imposes a limit on the CMT approach.

\begin{figure}[t]
\includegraphics[width=1\textwidth,trim={0.0cm 4cm 0cm 4cm},clip]{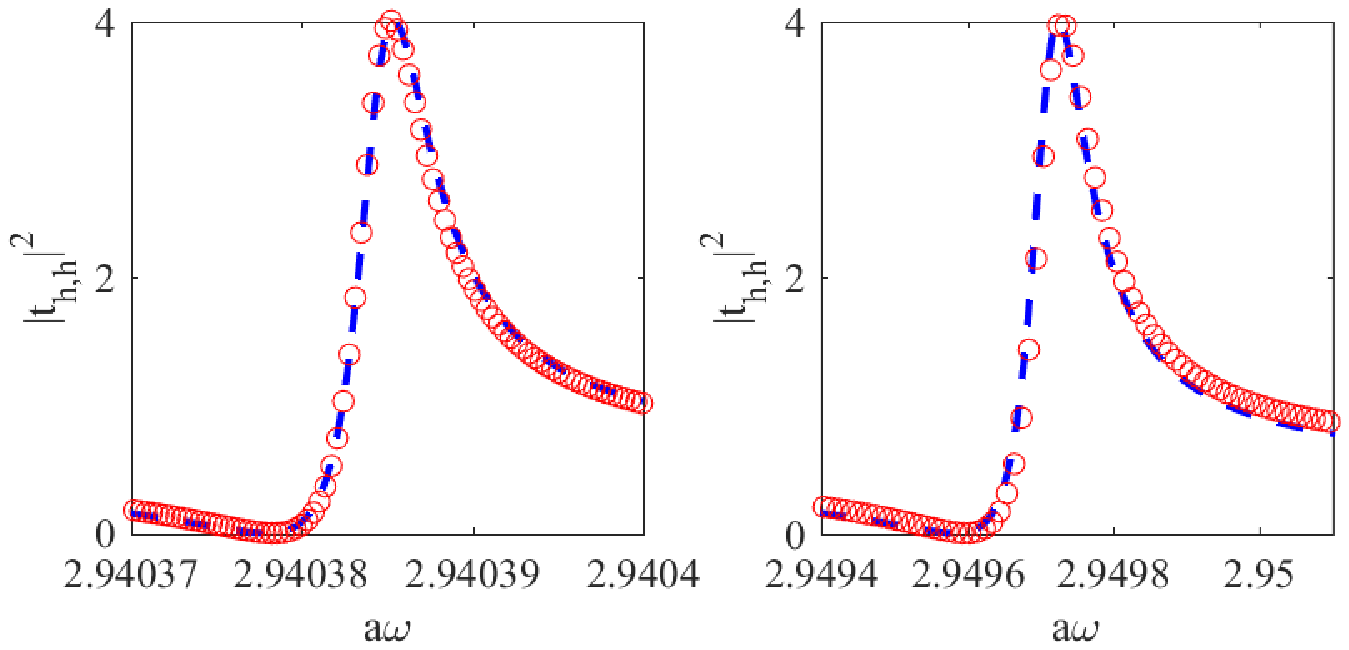}
\caption{(Color online) Optical response induced by the non-symmetry protected BIC from Fig. \ref{fig1} (b).
Blue dash line - exact numerical solution by Eq. (\ref{matrix}), red circles - CMT approximation. Left panel - $a\beta=0.1837$, right panel - $a\beta=0.3878$.  }
\label{fig5}
\end{figure}

Since the polarization conversion is forbidden for $m=0$ the scattered EM field is described by a single entry $t_{h,h}$ of matrix $\widehat{\cal T}$,
Eq. (\ref{Tdef}).
The only relevant element of the scattering matrix of the direct process
can be found from Eq. (\ref{cond1}) as
\begin{equation}\label{Cm0}
\{\widehat{\cal C}\}_{h,h}= {\gamma_h}^2/{|\gamma_h|^2}.
\end{equation}
By using Eq. (\ref{CMT_solution}) the final solution for $t_{h,h}$ can be written as
\begin{equation}\label{thh}
t_{h,h}=-1-e^{i2\chi}\frac{\omega-\omega_0-i\gamma}{\omega-\omega_0+i\gamma},
\end{equation}
where $\chi$ is the phase of the CMT coupling coefficient $\gamma_h$, $\chi=arg(\gamma_h)$. The dependance of $\chi$ on $\beta$
is shown in the inset to Fig. \ref{fig3} (b).
In Fig. \ref{fig4} we show the CMT spectrum Eq. (\ref{thh}) of the reflection coefficient $t_{h,h}$ near the $\Gamma$-point for the leaky zone with
the symmetry protected BIC in Fig. \ref{fig1} (a) in comparison with the exact numerical solution obtained from Eq. (\ref{matrix}). Two values of $\beta$
are chosen for numerical simulations to demonstrate that the accuracy of the CMT solution drops away off the $\Gamma$-point. However, one can see from
Fig. \ref{fig4} that even at the distance of one fourth of the Brillouin zone the CMT approach is capable of
reproducing the Fano feature in the reflection spectrum.

Similar simulations were undertaken for the non-symmetry protected BIC in Fig. \ref{fig1} (b). The results are shown in Fig. \ref{fig5}. Again one
can see a good agreement between the CMT and exact solutions far off the $\Gamma$-point. Although, in the present case the accuracy drops faster
than for the symmetry protected BIC.

\subsection{TM-modes}
\begin{figure}[t]
\includegraphics[width=1\textwidth,trim={0.0cm 4cm 0cm 4cm},clip]{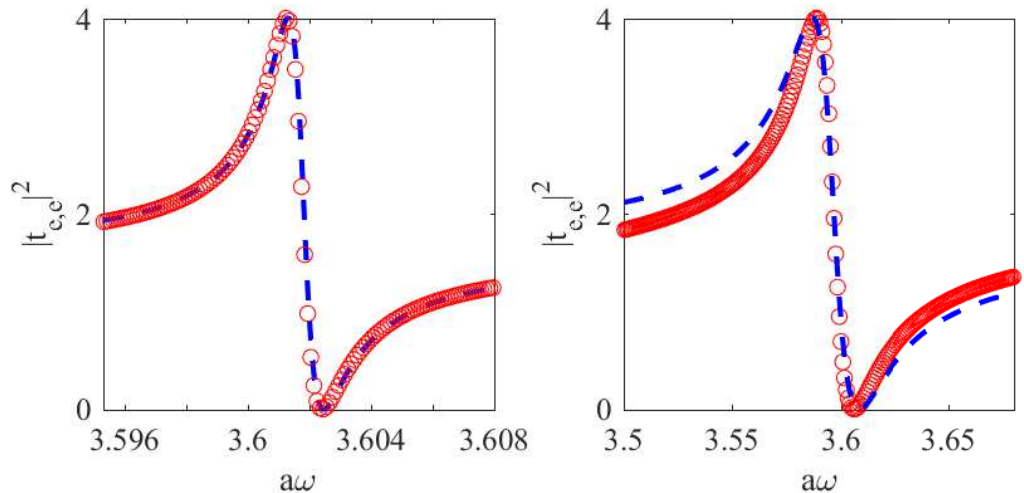}
\caption{(Color online) Optical response induced by symmetry protected TM BIC from Fig. \ref{fig2} (c).
Blue dash line - exact numerical solution by Eq. (\ref{matrix}), red circles - CMT approximation. Left panel - $a\beta=0.2041$,
right panel - $a\beta=0.7959$.  }
\label{fig6}
\end{figure}

CMT for TM-modes can be easily constructed along the same line as in the previous subsection. For vector ${\bf d}$ we have
\begin{equation}\label{d_vector_TM}
{\bf d}=
\left\{
\begin{array}{l}
\gamma_e \\
0
\end{array}
\right\}, \ 2\gamma=|\gamma_e|^2,
\end{equation}
while the coupling coefficient is found as
(see \ref{appendix}),
\begin{equation}\label{final1}
\gamma_e=\frac{\pm k_{\perp}}{2} \sqrt{\frac{1}{|\lambda_1| a k_0^3}}\overline{w}^{(e)}_{0}.
\end{equation}
The relevant entry of the scattering matrix of the direct process
reads
\begin{equation}\label{Cm0_TM}
\widehat{\cal C}_{e,e}=-{\gamma_e}^2/{|\gamma_e|^2}.
\end{equation}
In Fig. \ref{fig3}(d) we plot the resonant width found from Eq. (\ref{Cm0_TM}) in comparison against Eq. (\ref{Gamma}) to demonstrate that
the two approaches converge at the $\Gamma$-point.
The spectrum of the reflection coefficient $t_{e,e}$ for a symmetry protected TM BIC from Fig. \ref{fig2} (c) is shown in Fig. \ref{fig6} in comparison with
the CMT predictions. As before, we see a good agreement between the CMT and exact numerical solutions.

\section{Asymptotic behavior of the CMT coupling coefficients}\label{Sec5}

The expansion coefficients  ${a}^m_l, {b}^m_l$ for any quasi-mode have the following property as functions of the Bloch wave number $\beta$
\begin{eqnarray}
{a}^m_l(-\beta)=(-1)^{l+s}{a}^m_l(\beta), \notag \\ \label{ab_properties}
 {b}^m_l(-\beta)=(-1)^{l+1+s}{b}^m_l(\beta),
\end{eqnarray}
where $s=0,1$.
This property reflects the symmetry of the system defined by simultaneous change of the sign of $\beta$ and mirror reflection in the plane
perpendicular to the array axis $z$. In this section we consider the asymptotic behavior of the CMT coefficients against $\beta$
in the spectral vicinity of a BIC, i.e. at near normal incidence,
having in mind Eq. (\ref{ab_properties}).

\subsection{Symmetry protected BIC with $m=0$}
\begin{figure}[t]
\includegraphics[width=1\textwidth,trim={0.0cm 4cm 0cm 4cm},clip]{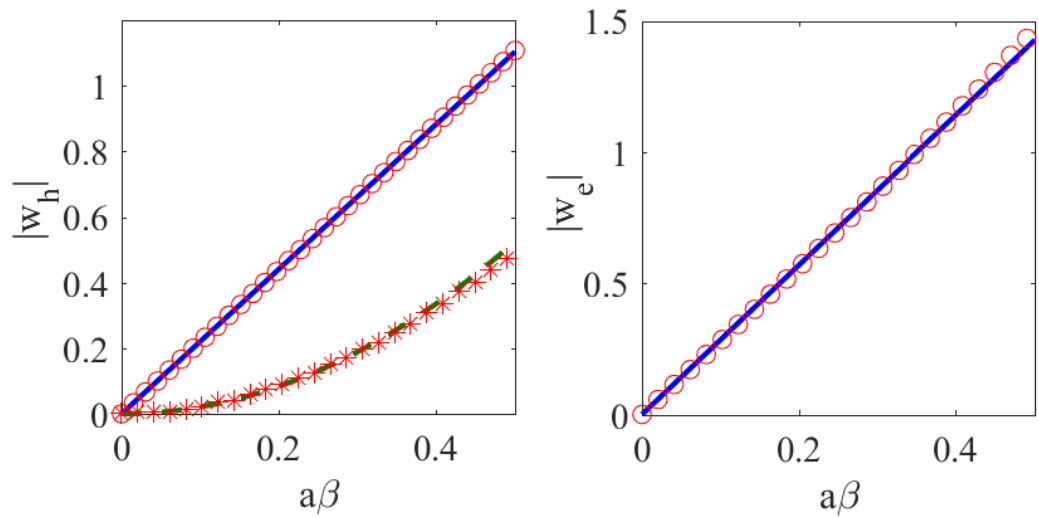}
\caption{(Color online) Left panel: quasi-mode coupling strength $w_h$ for the TE leaky modes Fig. \ref{fig2} (d)
The mode with a asymmetry protected BIC -
red circles; the mode with a non-symmetry protected BIC - red stars. The solid blue and dash green line show the best fit with linear
and parabolic functions, respectively. Right panel: Red circles show the quasi-mode coupling strength $w_e$ for the leaky mode
with symmetry protected TM BIC from Fig. \ref{fig2} (c). The solid blue line is the best fit with a linear
function.}
\label{fig7}
\end{figure}

Such in-$\Gamma$ BICs can be either TE-, or TM-polarized. Here we consider TE-polarized standing wave
BIC. This BIC is hosted by a TE-polarized leaky zone with the corresponding left eigenvector
written as
\begin{equation}\label{meq0}
{\bf y}=\left\{ {a}^0_{l}(\beta),  0\right\}.
\end{equation}
As far as the solution is symmetrically mismatched with TE scattering channel, we have $s=0$ in Eq. (\ref{ab_properties}).
It means that all odd expansion coefficients are zero in the $\Gamma$-point \cite{Bulgakov2015}, ${a}^m_{2k+1}(0)=0$.
Applying Eqs. (\ref{wh}, \ref{we}) and noticing that $\tau_{n,0}=0$ according to Eq. (\ref{pi_tau}) we have
\begin{eqnarray}\label{sym_m1}
w^{(h)}(\beta)=\sum_k F_{2k}^0\pi^0_{2k}(\theta){a}^0_{2k}(\beta)+\sum_k F_{2k+1}^0\pi^0_{2k+1}(\theta){a}^0_{2k+1}(\beta)\\
w^{(e)}(\beta)=0
\end{eqnarray}
By recollecting that $\cos(\theta)=\beta/k_0$ and using Eq. (\ref{pi_tau}), one can show that
\begin{eqnarray}
\tau_{l,m}(-\beta)=(-1)^{l+m}\tau_{l,m}(\beta), \notag \\
\pi_{l,m}(-\beta)=(-1)^{l+m+1}\pi_{l,m}(\beta). \label{pi_tau_properties}
\end{eqnarray}
Next refereing to  Eq. (\ref{ab_properties}) with $s=0$, we immediately see that both summands in Eq. (\ref{sym_m1}) are
odd with $\beta$. Hence the leading term expansion at the $\Gamma$-point is given by
\begin{equation}
w^{(h)}(\beta) \propto \beta.
\end{equation}
Finally, according to Eqs. (\ref{d_vector}, \ref{gamma})
the CMT coupling coefficient and resonant width have the following asymptotics
\begin{equation}
\gamma_h \propto \beta, \ \Gamma\propto \beta^2
\end{equation}
It is worth mentioning that the same arguments lead to the identical results for the symmetry protected TM-modes. The dependance of the
quasi-mode coupling strength against $\beta$ is shown in Fig. \ref{fig7} along with the best fit with a linear function.

\subsection{Non-symmetry protected BIC with $m=0$}
In that case the TE-polarized standing wave
BIC  is hosted by leaky zone with $s=1$ in Eq. (\ref{ab_properties}), and now
all even expansion coefficients are zero in the $\Gamma$-point \cite{Bulgakov2015}, ${a}^m_{2k}(\beta)=0$. Consequently, both summands in Eq. (\ref{sym_m1}) are even
with $\beta$. In general, at
the $\Gamma$-point we have
\begin{eqnarray}\label{asym2}
w^{(h)}(\beta) \propto  Const + \beta^2 \\
w^{(e)}(\beta)=0
\end{eqnarray}
Notice that $w^{(h)}(0)$ may be non-vanishing quantity and the leaky zone not necessarily has a BIC in the $\Gamma$-point.
That explains the difficulty in finding non-symmetry protected BICs in contrast
to symmetry protected ones. That always requires tuning the material parameters of system to eliminated the constant term in Eq. (\ref{asym2}).
However, such a BIC
is the most appealing to be employed for enhancement of light matter interaction due to higher order asymptotics of the CMT parameters
\begin{equation}\label{asym_CMT}
\gamma_h \propto \beta^2, \ \Gamma\propto \beta^4
\end{equation}
In Fig. \ref{fig7} we also show the asymptotic behavior of $w_e$ in the vicinity of the $\Gamma$-point which is well fit to a parabola.

\section{Conclusion}

We considered the optical response induced by bound states in the continuum in arrays of dielectric spheres.
By combining quasi-mode expansion technique with coupled mode theory (CMT) we put forward a theory of the optical response
by high-Q resonance surrounding BICs in momentum space. The central result are
analytical expressions for the CMT parameters, which can be easily calculated from the eigenfrequencies and eigenvectors
of the interaction matrix of the scattering systems. Once the CMT coupling coefficients are known
the optical response can be easily calculated through simple formulas. The results obtained are verified in comparison against exact
numerical solutions to demonstrate that the CMT approximation is capable of reproducing Fano features in the spectral vicinity of
the BIC. We expect that the proposed approach is not limited to arrays of dielectric sphere. The complicated machinery related
to cylindrical and spherical wave expansions is necessitated by our choice of the computational method. However, if the scattering problem
is cast in form of a set of liner equations similar to Eq. (\ref{matrix}) by means of, say, finite difference or finite element methods,
the quasi-mode expansion technique becomes equally applicable for any periodic structure supporting bound states in the continuum. Thus, we speculate
that our results may be useful for engineering the resonant response in systems with diverging Q-factor.

So far the asymptotic behavior of the critical light enhancement induced by BICs has been rigorously studied only in perforated slabs \cite{Shipman2005},
and arrays of dielectric rods \cite{Yuan17,Yuan18}. In this paper we put focus on linear arrays of dielectric spheres.
Based on the quasi-mode expansion technique we derived the asymptotic scaling law for the CMT parameters in the vicinity of
the $\Gamma$-point. It is rigourously demonstrated that the line width in the CMT approximation exhibits different asymptotic behavior
depending on the symmetry of the BIC. In particular, it is proved that for symmetry protected BIC the Q-factor diverges as $\beta^2$ in the vicinity
of the $\Gamma$-point. At the same time, for the BICs unprotected by symmetry the Q-factor diverges as $\beta^4$. That suggests application
of the non-symmetry protected BICs to achieve a stronger resonant effect for enhancement of light-matter interaction. We mention in passing
that our findings totally comply with earlier observations \cite{Bulgakov17,Yuan17,Yuan18}.

In this paper the analysis was limited to the case of zero angular momentum, i.e. to the situation when the waves of TE and TM polarization are
decoupled. It would be interesting, though, to extend the results to the BIC with non-zero OAM \cite{Bulgakov17a},
in which case the leaky mode are hybridized and matrix $\widehat{\cal T}$ is no more diagonal. This brings up certain difficulties in constructing
CMT, since the matrix of the direct process is no more uniquely defined by the quasi-mode coupling strengths. Moreover, an extra difficulty arises
in defining the phase of the CMT coupling constant since Eq. (\ref{left_right}) is not applicable in subspaces with non-zero orbital angular momentum.
First principle derivation of CMT from the quasi-modal expansion for waves with non-zero OAM is our goal for future studies.

\section*{Acknowledgment} This work was supported by Ministry of Education and Science of Russian Federation
(contract N 3.1845.2017) and RFBR grant 16-02-00314.

\section*{References}
\bibliographystyle{unsrt}
\bibliography{BSC_light_trapping}

\appendix
\section{} \label{appendix}
The solution for EM-field for TE-quasi mode {\bf x} with $m=0$ can be written as
\begin{equation}\label{E}
{\bf E}=\sum_je^{ija\beta}\sum_{l=m^*}^{\infty}\tilde{a}_l^0{\bf M}_l^0({\bf r}_j),
\end{equation}
where $\tilde{a}_l^0$ are the expansion coefficients for the right eigenvector ${\bf x}$
\begin{equation}
{\bf x}=\left\{\tilde{a}_l^0\right\}.
\end{equation}
On the other hand spherical vector harmonics ${\bf M}_l^0({\bf r}_j)$ can be expressed through
a scalar function $\psi_l^{0}({\bf r}_j)$ as \cite{Stratton41}
\begin{equation}\label{M}
{\bf M}_l^0({\bf r}_j)=\nabla\times[{\bf r}_j\psi_l^{0}({\bf r}_j)],
\end{equation}
where
\begin{equation}
\psi_l^{0}({\bf r}_j)=\frac{1}{\sqrt{l(l+1)}}h^{(1)}_l(k_0{\bf r}_j)Y^0_l(\theta,\phi)
\end{equation}
with $h^{(1)}_l(x)$ as the spherical Hankel function, and
\begin{equation}
Y^m_l(\theta,\phi)=(-1)^m\sqrt{\frac{(2l+1)(l-m)!}{4\pi(l+m)!}}e^{im\phi}P^{m}_l[\cos(\theta)].
\end{equation}
Combining Eqs. (\ref{E}, \ref{M}) for the azimuthal component of the electric field $E_{\phi}({\bf r})$ we have
\begin{equation}\label{Ephi}
E_{\phi}({\bf r})=-e^{-i\phi}\sum_je^{ija\beta}\sum^{\infty}_{l=1}\tilde{a}^{0}_lh^{(1)}_l(k_0{\bf r}_j)Y^1_l(\theta,\phi),
\end{equation}
where we used
\begin{equation}
\frac{\partial P^{0}_l[\cos(\theta)]}{\partial \theta}=-P^{1}_l[\cos(\theta)]
\end{equation}
According to \cite{Thompson10} the series (\ref{Ephi}) can be rewritten as an expansion into cylindrical functions
\begin{equation}
E_{\phi}({\bf r})=-\sum_{n=-\infty}^{\infty}M_ne^{i\beta_nz}K_1\left(k_0 \rho \sqrt{\left(\frac{\beta_n}{k_0} \right)^2-1} \right),
\end{equation}
where $K_1(x)$ is modified Bessel function,
\begin{equation}
\beta_n=\beta+\frac{2\pi}{a}n,
\end{equation}
and
\begin{equation}\label{Mn}
M_n=\frac{2i}{k_0a}\sum_{l=1}^{\infty}i^{-l}\tilde{a}_l^0\sqrt{\frac{(2l+1)}{4\pi l (l+1)}}P_l^1\left(\frac{\beta_n}{k_0} \right).
\end{equation}

In the next step we recall that under condition Eq. (\ref{domain}) only the zeroth diffraction order $n=0$ is present in the far-field radiation.
In the far field all waves can be written in the form of expansion into cylindrical functions.
For such cylindrical waves the electric $E_{\phi}({\bf r})$ is linked to the axial component of
the magnetic field $H_{z}({\bf r})$ through the following equation
\begin{equation}
E_{\phi}({\bf r})=-\frac{ik_0}{k_{\perp}^2}\frac{\partial H_{z}({\bf r})}{\partial \rho}.
\end{equation}
By using
\begin{equation}
K_1(-ix)=-\frac{\pi}{2} H_1^{(1)}(x),
\end{equation}
and
\begin{equation}
H_1^{(1)}(x)=-\frac{\partial H_0^{(1)}(x)}{\partial x}
\end{equation}
we obtain the far-zone EM field pattern generated by the optical quasi-mode
\begin{equation}\label{H_z}
H_z({\bf r})=-\frac{i\pi}{2}\frac{k_{\perp}}{k_0}M_0e^{i\beta z} H_0^{(1)}(k_{\perp}\rho).
\end{equation}
The eigenvectors ${\bf y}$, ${\bf x}$ are defined up to an arbitrary phase factor.
Since the interaction matrix possesses the property
$\{\widehat{\cal L}\}_{l,l'}=(-1)^{l+l'}\{\widehat{\cal L}\}_{l',l}$ one can show that vector $\overline{\bf y}$,
\begin{equation}\label{left_right}
\{\overline{\bf y}\}_l=(-1)^l\{{\bf x}^*\}_l
\end{equation}
is a left eigenvector of $\widehat{{\cal L}}$. In what follows we set
\begin{equation}
{\bf y}_n=\overline{\bf y}_n.
\end{equation}
which allow us to eliminate the disambiquety of the phase definition \cite{Bulgakov17}.
Using the definition of the expansion coefficient for vector ${\bf q}$, Eq. (\ref{cd_def}) we find
\begin{equation}
M_0=\frac{2i\sqrt{C({\bf k})}k_{\perp}}{k_0^2a}\frac{1}{4\pi}{\bf q}^{\dag} {\bf x}.
\end{equation}
Applying Eq. (\ref{left_right}) according to \cite{Bulgakov17} one can show that ${\bf q}^{\dag}{\bf x}={\bf y}^{\dag}{\bf q}$,  which together with
Eq. (\ref{wh}) yields
\begin{equation}
M_0=\frac{2i\sqrt{C({\bf k})}k_{\perp}}{k_0^2a}\frac{1}{4\pi}{w}^{(h)}.
\end{equation}
Physically, Eq. (\ref{H_z}) is the far-field pattern of the quasi-mode specified by vector ${\bf x}$, therefore
the far-field solution produced by the term Eq. (\ref{resonant}) with vanishing $\lambda$ in denominator can be written as
\begin{equation}
H^{({\rm out})}_z({\bf r})=\frac{k_{\perp}^2\sqrt{C({\bf k})}}{4ak_0^3}\frac{\left(w^{(h)}\right)^2}{\lambda}H_0^{(1)}(k_{\perp}\rho)e^{i\beta z}a^{(+)}_h.
\end{equation}
Using the series expansion Eq.(\ref{exp_def}) we find the resonant contribution into the outgoing far field radiation as
\begin{equation}\label{res1}
H^{\rm (res)}_z({\bf r})=\frac{k_{\perp}^2\sqrt{C({\bf k})}}{i4\lambda_1ak_0^3}\frac{\left(w^{(h)}_{0}\right)^2}{i(\omega_0-\omega)+\gamma}H_0^{(1)}(k_{\perp}\rho)e^{i\beta z}a^{(+)}_h.
\end{equation}
On the other hand according to Eqs. (\ref{CMT_solution}, \ref{d_vector}) the CMT solution for the resonant term reads
\begin{equation}\label{res2}
H^{\rm (res)}_z({\bf r})=-\gamma_h^2\frac{1}{\sqrt{C({\bf k})}}
\frac{1}{i(\omega_0-\omega)+\gamma}H_0^{(1)}(k_{\perp}\rho)e^{i\beta z}a^{(+)}_h.
\end{equation}
Comparing Eq. (\ref{res1}) against Eq. (\ref{res2}) we find
\begin{equation}\label{final_final}
\gamma_h^2=-\frac{k_{\perp}^2C({\bf k})}{i4\lambda_1ak_0^3}\left(w^{(h)}_{0}\right)^2.
\end{equation}

According to Eq. (\ref{wh}) the quantity $w^{(h)}_{0}$ is introduced with $\sqrt{C({\bf k})}$ in the denominator. One can see that
the final result is independent of $C({\bf k})$ which
can be removed from Eq. (\ref{final_final}). By using
\begin{equation}
\overline{w}^{(h)}_{0}={w}^{(h)}_{0}\sqrt{C({\bf k})}
\end{equation}
that yields
\begin{equation}
\gamma_h^2=-\frac{k_{\perp}^2}{i4\lambda_1ak_0^3}\left(\overline{w}^{(h)}_{0}\right)^2.
\end{equation}
Finally, by recollecting that $\lambda_1$ is imaginary positive for $\gamma_h$ we have Eq. (\ref{final}). Essentially the
same derivation can be equally applied to the TM-modes with $m=0$ leading to Eq. (\ref{final1}) with
\begin{equation}
\overline{w}^{(e)}_{0}={w}^{(e)}_{0}\sqrt{C({\bf k})}.
\end{equation}

\end{document}